\documentclass[numbers]{article}

\usepackage[numbers]{natbib}

\usepackage{arxiv}

\usepackage[utf8]{inputenc} 
\usepackage[T1]{fontenc}    
\usepackage{hyperref}       
\usepackage{url}            
\usepackage{booktabs}       
\usepackage{amsfonts}       
\usepackage{nicefrac}       
\usepackage{microtype}      
\usepackage{lipsum}		
\usepackage{graphicx}
\usepackage{natbib}
\usepackage{doi}
\usepackage{amsmath,mathtools}
\usepackage{amssymb}
\usepackage{algorithm}
\usepackage[noend]{algpseudocode}

\newcommand*\Let[2]{\State #1 $\gets$ #2}
\algrenewcommand\alglinenumber[1]{
    {\sf\footnotesize#1}}
\algrenewcommand\algorithmicrequire{\textbf{Precondition:}}
\algrenewcommand\algorithmicensure{\textbf{Postcondition:}}

\DeclareMathOperator{\diag}{diag}

\title{Iterated and exponentially weighted moving \\ principal component analysis}

\author{ \href{https://orcid.org/0000-0001-6846-6649}{\includegraphics[scale=0.06]{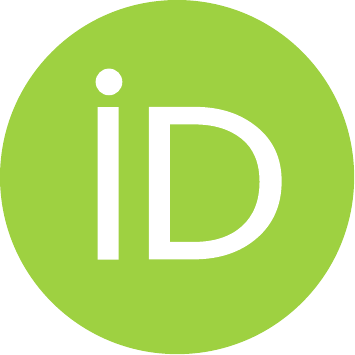}\hspace{1mm}Paul Bilokon} \\
	Departments of Computing and Mathematics \\
	Imperial College London \\
	South Kensington Campus \\
    London SW7 2AZ, UK \\
	\texttt{paul.bilokon@imperial.ac.uk} \\
	\And
	\href{https://orcid.org/0000-0000-0000-0000}{\includegraphics[scale=0.06]{orcid.pdf}\hspace{1mm}David Finkelstein} \\
	DQR Ltd \\
	3rd Floor, 120 Baker Street \\
	London W1U 6TU, UK \\
	\texttt{df@dqr.am} \\
}

\renewcommand{\shorttitle}{IPCA and EWMPCA}

\hypersetup{
pdftitle={Iterated and exponentially weighted moving principal component analysis},
pdfsubject={q-fin.ST},
pdfauthor={Paul Bilokon, David Finkelstein},
pdfkeywords={principal component analysis, PCA, moving statistics, rolling statistics},
}

\begin{document}
\maketitle

\begin{abstract}
The principal component analysis~(PCA) is a staple statistical and unsupervised machine learning technique in finance. The application of PCA in a financial setting is associated with several technical difficulties, such as numerical instability and nonstationarity. We attempt to resolve them by proposing two new variants of PCA: an iterated principal component analysis (IPCA) and an exponentially weighted moving principal component analysis (EWMPCA). Both variants rely on the Ogita--Aishima iteration as a crucial step.
\end{abstract}

\keywords{principal component analysis \and PCA \and moving statistics \and rolling statistics}

\section{Introduction}

The \emph{principal component analysis~(PCA)}~\cite{jolliffe-2002, jolliffe-2016} invented by Pearson~\cite{pearson-1901} and improved by Hotelling~\cite{hotelling-1933, hotelling-1936} is a staple statistical and unsupervised machine learning technique in finance~\cite{aldridge-2021}. Its central idea is to reduce the dimensionality of a data set consisting of a large number of interrelated variables, while retaining as much as possible of the variation present in the data set. This is achieved by transforming to a new set of variables, the \emph{principal components (PCs)}, which are uncorrelated, and which are ordered so that the first few retain most of the variation present in all of the original variables.

\begin{figure}
\centering
\includegraphics[width=1.\textwidth]{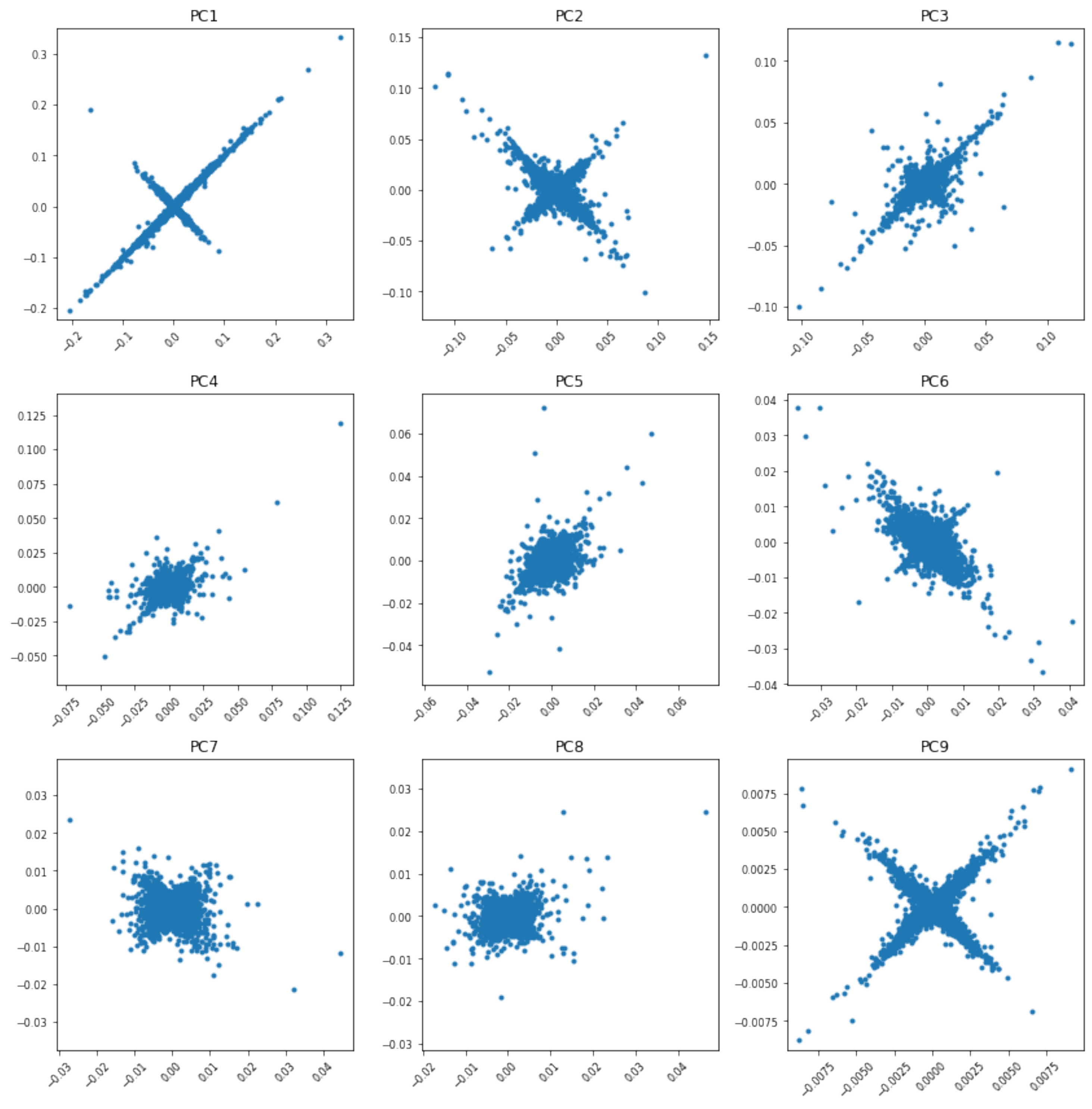}
\caption{We perform classical PCA on Data Set~1 twice. In the first instance, we perform PCA on the entire dataset. In the second, we perform PCA on each year individually and stack the results together. We then produce scatter plots, where the second result ($y$-axis) is plotted against the first result ($x$-axis).}
\label{fig:1}
\end{figure}

The application of PCA in a financial setting is associated with several technical difficulties. First, the entire data set may not be immediately available (it may be arriving piecewise in real time), so one is forced to work with its subsets pertaining to different time intervals. When the PCs are computed separately on each subset, the geometry of the resulting PCs may suffer from numerical artifacts, as illustrated in Figure~\ref{fig:1}. In particular, the sign of a given PC may ``flip'' from one subset to the next. Second, financial data are rarely stationary, and the assumption of a constant covariance matrix is rarely justified.

To remedy the first problem, we propose an \emph{iterated principal component analysis~(IPCA)}: instead of computing the principal components on each arriving subset independently, we iteratively refine them from one subset to the next. To remedy the second problem, we combine the aforementioned iterative refinement with an exponentially weighted moving computation of the covariance matrix, to obtain an \emph{exponentially weighted moving principal component analysis~(EWMPCA)}.

We are heavily indebted to Ogita and Aishima, who proposed an iterative refinement method for symmetric eigenvalue decomposition~\cite{ogita-2018}, on whose work we build.

We have used two data sets in this study. Both are derived from data supplied by FirstRate Data and both consist of hourly returns on futures. The first data set (Data Set 1) covers the period 20th August, 2007 to 4th June, 2021, both inclusive, and consists of hourly returns on equity futures: DAX~(DY), E-Mini S\&P~500~(ES), E-Mini S\&P~500 Midcap~(EW), Euro Stoxx~50~(FX), CAC40~(MX), E-Mini Nasdaq-100~(NQ), E-Mini Russell~2000~(RTY), FTSE~100~(X), Dow Mini~(YM). The second data set (Data Set 2) covers the period 10th September, 2012 to 4th June, 2021, both inclusive, and consists of hourly returns on fuel futures: Brent Last Day Financial~(BZ), Crude Oil WTI~(CL), Natural Gas (Henry Hub) Last-day Financial~(HH), NY Harbor ULSD (Heating Oil)~(HO), Henry Hub Natural Gas~(NG), RBOB Gasoline~(RB).

\section{Classical PCA}

A data set of $n$ observations of $p$ features can be represented by an $n \times p$ data matrix $X$, whose $j$th column, $x_{:,j}$, is the vector of $n$ observations of the $j$th feature. We seek a linear combination of the columns of matrix $X$ with maximum variance. Such linear combinations are given by $X w^{(1)}$, where $w^{(1)}$ is a vector of constants $w^{(1)}_1, \ldots, w^{(1)}_p$. It can be shown that $w^{(1)}$ must be a (unit-norm) eigenvector of the sample covariance matrix $Q$ associated with the data set; more precisely, the eigenvector corresponding to the largest eigenvalue $\lambda^{(1)}$ of $Q$.

The full set of eigenvectors $w^{(1)}, \ldots, w^{(p)}$ of $Q$, corresponding to the eigenvalues sorted in decreasing order, $\lambda^{(1)}, \ldots, \lambda^{(p)}$, are the solutions to the problem of obtaining up to $p$ linear combinations $X w^{(k)}$, $1 \leq k \leq p$, which successively maximize variance, subject to uncorrelatedness with previous linear combinations. We call these linear combinations the \emph{principal components~(PCs)} of the data set.

It is standard to define PCs as the linear combinations of the \emph{centred} variables $x^*_{:,j}$, with generic element $x^*_{ij} = x_{ij} - \bar{x}_{:,j}$, where $\bar{x}_{:,j}$ denotes the mean value of the observations on variable $j$.

At the core of PCA is the eigendecomposition of the sample covariance matrix $Q$ or, equivalently, the singular value decomposition~(SVD) of the data matrix $X$.

An industry-standard implementation of PCA is \texttt{sklearn.decomposition.PCA} in the software library \textsf{scikit-learn}~\cite{pedregosa-2011}. It uses the LAPACK~\cite{anderson-1999} implementation of the full SVD or a randomized truncated SVD by the method of Halko \textit{et al.}~\cite{halko-2011}. When comparing our results to the classical PCA it is this implementation that we use as a benchmark.

\section{The Ogita--Aishima algorithm}

As is easy to see, Algorithm~\ref{alg:estimate-eigenvalues} estimates the eigenvalues of a given real symmetric matrix $A \in \mathbb{R}^{n \times n}$ for a precomputed set of eigenvectors $\hat{X} \in \mathbb{R}^{n \times n}$ (the eigenvectors are in the columns of $\hat{X}$).
\begin{algorithm}
  \caption{Estimate the eigenvalues of a given real symmetric matrix $A$ corresponding to the precomputed eigenvectors. \label{alg:estimate-eigenvalues}}
  \begin{algorithmic}[1]
    \Statex
    \Function{estimate\_eigenvalues}{$A$, $\hat{X}$, return\_extra}
      \Let{$R$}{$I - \hat{X}^{\intercal} \hat{X}$}
      \Let{$S$}{$\hat{X}^{\intercal} A \hat{X}$}
      \For{$i \gets 1 \textrm{ to } n$}
        \Let{$\tilde{\lambda}_i$}{$s_{ii} / (1 - r_{ii})$}
      \EndFor
      \If{return\_extra}
          \State \Return{$\tilde{\lambda}$, $R$, $S$}
      \Else
          \State \Return{$\tilde{\lambda}$}
      \EndIf
    \EndFunction
  \end{algorithmic}
\end{algorithm}

Ogita and Aishima have proposed and analyzed an iterative refinement algorithm~\cite[Algorithm~1]{ogita-2018} for approximate eigenvectors $\hat{X}$ of $A$. We list this algorithm here as Algorithm~\ref{alg:ogita-aishima-step}. The authors demonstrate the monotone and quadratic convergence of the algorithm under some reasonable technical conditions.

\begin{algorithm}
  \caption{Refinement of approximate eigenvectors of a real symmetric matrix. \label{alg:ogita-aishima-step}}
  \begin{algorithmic}[1]
    \Statex
    \Function{ogita\_aishima\_step}{$A, \hat{X}$}
      \Let{$\tilde{\lambda}$, $R$, $S$}{\Call{estimate\_eigenvalues}{$A$, $\hat{X}$, \textbf{true}}} \Comment{Compute approximate eigenvalues.}
      \Let{$\tilde{D}$}{$\diag(\tilde{\lambda}_i)$}
      \Let{$\delta$}{$2(\|S - \tilde{D}\|_2 + \|A\|_2 \|R\|_2)$}
      \For{$i \gets 1 \textrm{ to } n$}
        \For{$j \gets 1 \textrm{ to } n$}
        \Let{$\tilde{e}_{ij}$}{$\left\{
                                  \begin{array}{ll}
                                    \frac{s_{ij} + \tilde{\lambda}_j r_{ij}}{\tilde{\lambda}_j - \tilde{\lambda}_i}, & \hbox{if $|\tilde{\lambda}_i - \tilde{\lambda}_j| > \delta$;} \\
                                    r_{ij}/2, & \hbox{otherwise.}
                                  \end{array}
                                \right.
        $} \Comment{Compute $\tilde{E}$.}
        \EndFor
      \EndFor
      \Let{$X'$}{$\hat{X} + \hat{X} \tilde{E}$}
      \State \Return{$X'$}
    \EndFunction
  \end{algorithmic}
\end{algorithm}

We wrap the function \texttt{ogita\_aishima\_step} in a higher-level function that performs the number of iterations required for satisfying a sensible convergence criterion (Algorithm~\ref{alg:ogita-aishima}).

\begin{algorithm}
  \caption{Take multiple Ogita--Aishima steps to achieve the required level of convergence. \label{alg:ogita-aishima}}
  \begin{algorithmic}[1]
    \Statex
    \Function{ogita\_aishima}{$A$, $\hat{X}$, tol=1e-6, max\_iter\_count=\textbf{none}, sort\_by\_eigenvalues=\textbf{false}}
      \Let{iter\_count}{0}
      \While{\textbf{true}}
        \Let{iter\_count}{iter\_count + 1}
        \Let{$\hat{X}'$}{\Call{ogita\_aishima\_step}{$A$, $\hat{X}$}}
        \If{max\_iter\_count \textbf{is not none and} iter\_count == max\_iter\_count}
          \State \textbf{break}
        \EndIf
        \Let{$\epsilon$}{$\|\hat{X}' - \hat{X}\|_2$}
        \If{$\epsilon$ < tol}
          \State \textbf{break}
        \EndIf
        \Let{$\hat{X}$}{$\hat{X}'$}
      \EndWhile
      \If{sort\_by\_eigenvalues}
        \Let{$\tilde{\lambda}$}{\Call{estimate\_eigenvalues}{$A$, $\hat{X}'$, \textbf{false}}}
        \State Sort $\tilde{\lambda}$ in descending order and reorder the corresponding columns of $\hat{X}'$ to match that order
      \EndIf
      \Return{$\hat{X}'$}
    \EndFunction
  \end{algorithmic}
\end{algorithm}

\section{Iterated PCA}

\emph{Iterated PCA~(IPCA)} is a straightforward extension of PCA wherein the algorithm can be fitted multiple times. Every time fit is invoked on a new data subset, that subset's sample covariance matrix $Q$ is calculated. The eigenvectors $\hat{W}$ of $Q$ are stored between the fits; for each new $Q$ the previous eigenvectors are used as an initial guess in \texttt{ogita\_aishima}($Q$, $\hat{W}$, \texttt{sort\_by\_eigenvalues}=\textbf{true}). (At the beginning, when no initial guess is available, the eigenvectors are obtained using standard methods.)

\begin{figure}
\centering
\includegraphics[width=1.\textwidth]{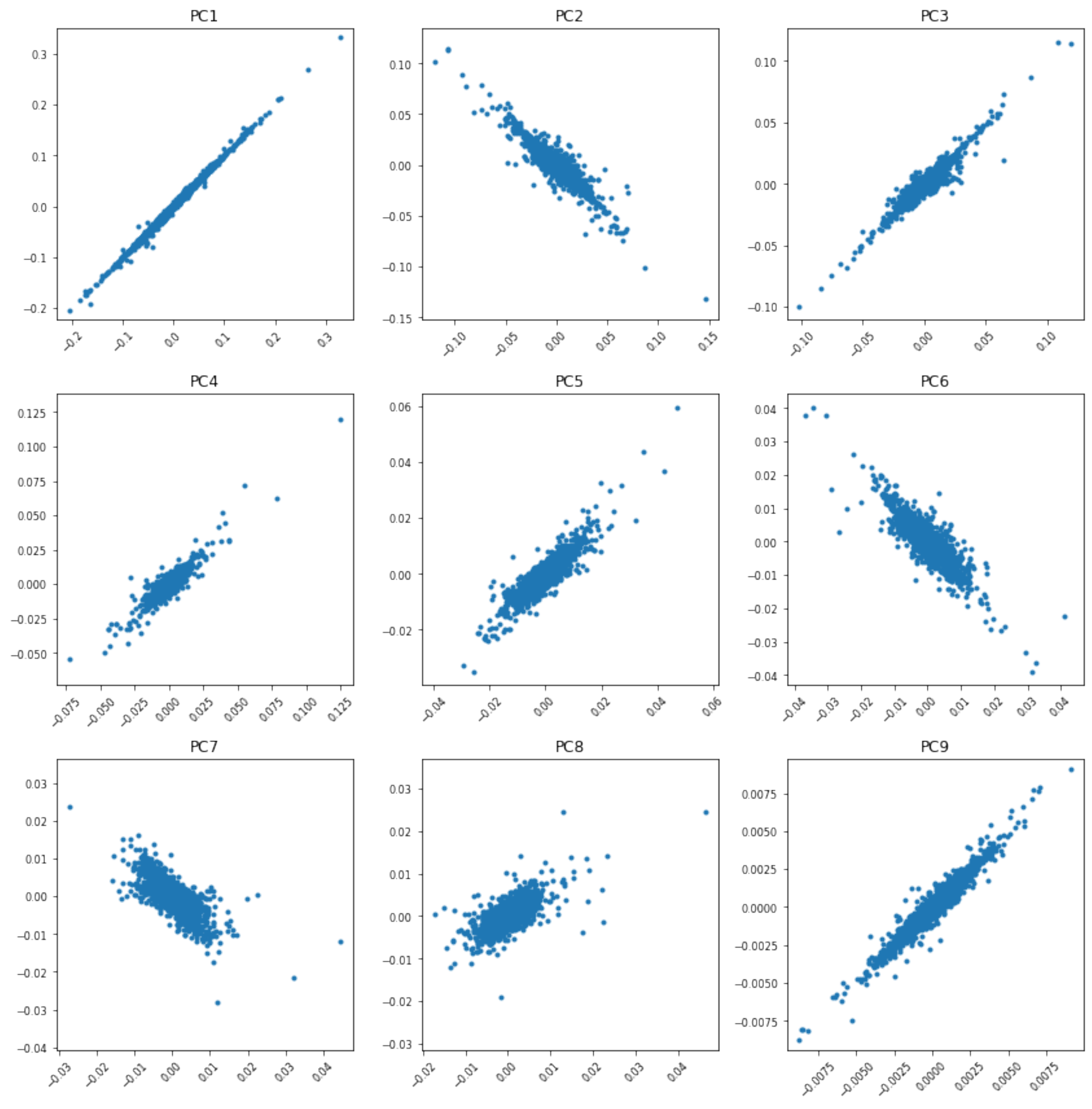}
\caption{We perform classical PCA on Data Set~1. Then we perform IPCA on each year individually and stack the results together. We then produce scatter plots, where the second result ($y$-axis) is plotted against the first result ($x$-axis).}
\label{fig:2}
\end{figure}

IPCA resolves the numerical instability problem witnessed in Figure~\ref{fig:1} as demonstrated by Figure~\ref{fig:2}.

\section{Moving statistics}

Let $x_1, \ldots, x_t$, $t \in \mathbb{N}$, be a sequence of $p$-dimensional observations. The \emph{exponentially weighted moving average} for this sequence can be calculated recursively as
\begin{equation*}
m_t = \left\{
        \begin{array}{ll}
          x_1, & \hbox{$t = 1$,} \\
          (1 - \alpha) x_t + \alpha m_{t-1}, & \hbox{$t > 1$,}
        \end{array}
      \right.
\end{equation*}
where $0 < \alpha < 1$ is a constant parameter.

Tsai~\cite{tsay-2010} proposes a similar moving statistic for the sample covariance:
\begin{equation*}
S_t = \left\{
        \begin{array}{ll}
          0_{p \times p}, & \hbox{$t = 1$,} \\
          (1 - \alpha) (x_t - m_t) (x_t - m_t)^{\intercal} + \alpha S_{t-1}, & \hbox{$t > 1$,}
        \end{array}
      \right.
\end{equation*}

One way to estimate the parameter $\alpha$ is by using maximum likelihood~(ML). For example, if $x_1, \ldots, x_T$, $t \in \mathbb{N}$, are normally distributed, then $\alpha_{ML}$ is the value of $\alpha$ that maximizes
\begin{equation*}
\ln \mathcal{L}(\alpha) \propto -\frac{1}{2} \sum_{t=1}^T |S_t| - \frac{1}{2} \sum_{t=1}^T (x_t - m_t)^{\intercal} S_t^{-1} (x_t - m_t).
\end{equation*}

In an example in Section~10.1 of~\cite{tsay-2010}, Tsai describes the value $\alpha \approx 0.9305$ as being in the typical range commonly seen in practice.

\begin{figure}
\centering
\includegraphics[width=1.\textwidth]{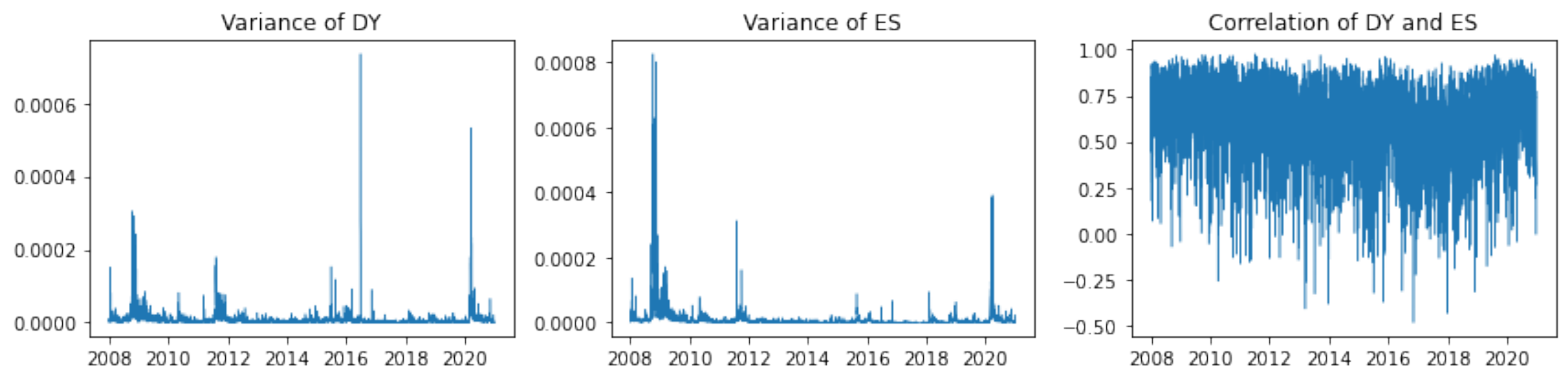}
\caption{The exponentially weighted moving covariance reveals the nonstationary nature of financial data.}
\label{fig:3}
\end{figure}

Moving statistics reveal the nonstationary nature of financial data. Consider Data Set~1 as an example. The variances of the returns on individual futures change over time and exhibit the so-called volatility clustering~\cite{cont-2007}; the correlations between pairs of futures are also time-varying (Figure~\ref{fig:3}).

\begin{figure}
\centering
\includegraphics[width=1.\textwidth]{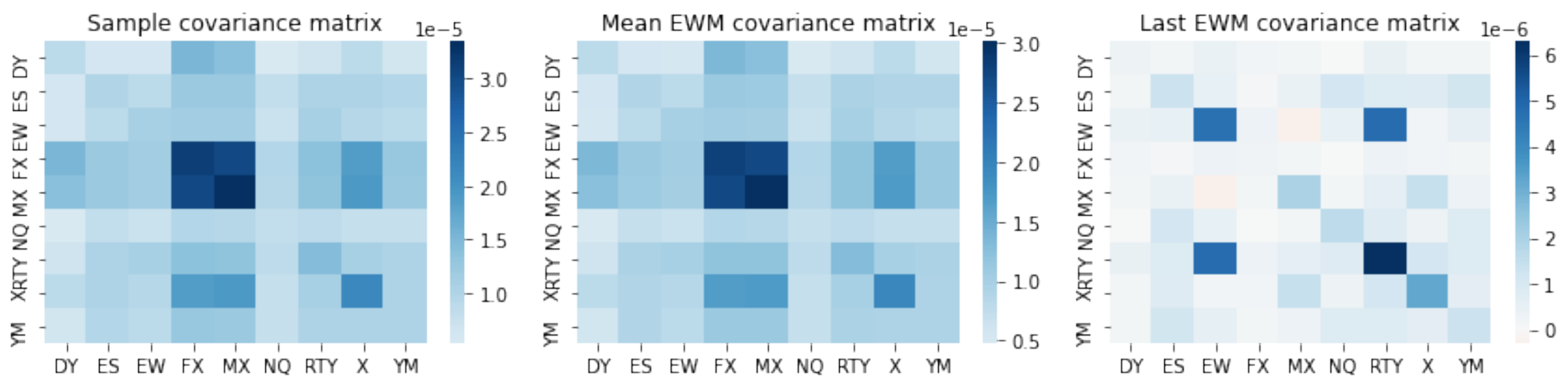}
\caption{Heatmaps comparing the sample covariance matrix with a time series of exponentially weighted covariance matrices.}
\label{fig:4}
\end{figure}

Whereas the \emph{mean} exponentially weighted moving covariance matrix resembles the sample covariance matrix, individual exponentially weighted moving covariance matrices (such as the last one in our time series shown in Figure~\ref{fig:4}) may differ from it significantly.

The principal components obtained using the sample covariance matrix present an averaged picture; we need a more precise tool to work out what's going on at each time step.

\section{Exponentially weighted moving PCA}

Combining ideas from the Ogita--Aishima iteration and moving statistics it is straightforward to formulate an \emph{exponentially weighted moving PCA~(EWMPCA)}---Algorithm~\ref{alg:ewmpca}.

\begin{algorithm}
  \caption{Exponentially weighted moving PCA. Here $X$ is the $n \times p$ data matrix, $\hat{W}^{\text{initial}}$ is the initial guess for the eigenvector matrix. \label{alg:ewmpca}}
  \begin{algorithmic}[1]
    \Statex
    \Function{ewmpca}{$X$, $\alpha$, $\hat{W}^{\text{initial}}$, tol=1e-6, max\_iter\_count=\textbf{none}}
      \For{$i = 1, \ldots, n$}
        \If{$i == 1$}
          \Let{$m$}{$x_{i,:}^{\intercal}$} \Comment{Exponentially weighted moving average.}
          \Let{$S$}{$0_{p \times p}$} \Comment{Exponentially weighted moving covariance.}
          \Let{$\hat{W}$}{$\hat{W}^{\text{initial}}$} \Comment{Eigenvectors of $S$.}
          \Let{$z_{i,:}$}{$0_{1 \times p}$} \Comment{Principal components.}
        \Else
          \Let{$m$}{$(1 - \alpha) x_{i,:}^{\intercal} + \alpha m$}
          \Let{$x^*$}{$x_{i,:}^{\intercal} - m$}
          \Let{$S$}{$(1 - \alpha) x^* (x^*)^{\intercal} + \alpha S$}
          \Let{$\hat{W}$}{\Call{ogita\_aishima}{$S$, $\hat{W}$, tol, max\_iter\_count, sort\_by\_eigenvalues=\textbf{true}}}
          \Let{$z_{i,:}$}{$(x^*)^{\intercal} \hat{W}$}
        \EndIf
      \EndFor
      \Return{$z$}
    \EndFunction
  \end{algorithmic}
\end{algorithm}

$\hat{W}^{\text{initial}}$ must be such as to facilitate convergence. One option is to use the sample covariance matrix for the first few (say 100) observations.

\begin{figure}
\centering
\includegraphics[width=1.\textwidth]{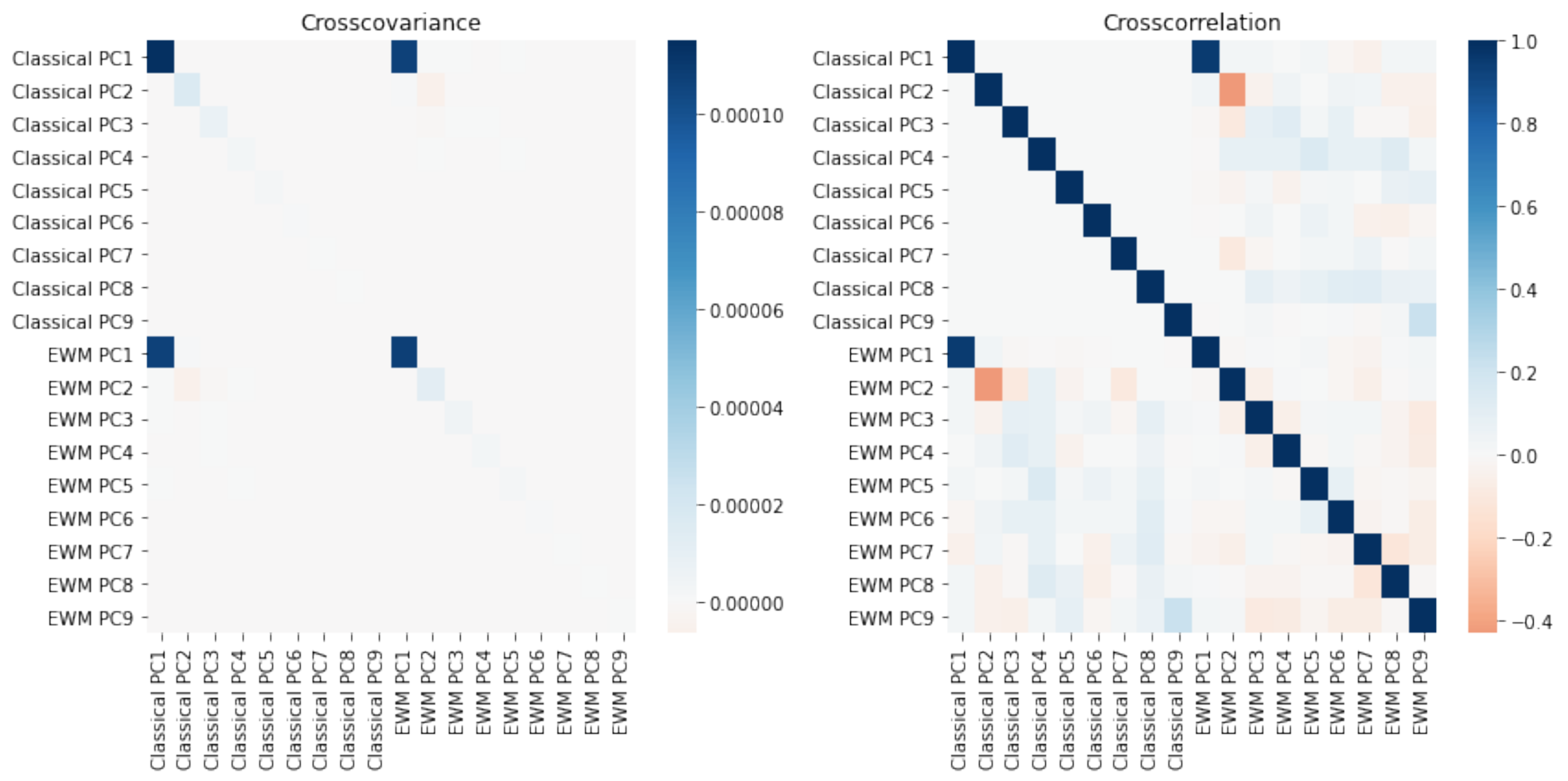}
\caption{Heatmaps of the crosscovariance and crosscorrelation between the classical PCA and EWMPCA principal components computed on Data Set~1.}
\label{fig:5}
\end{figure}

As we can see from Figure~\ref{fig:5}, the EWMPCA principal components are not pairwise uncorrelated; by construction, they are uncorrelated \emph{locally}, not \emph{on average}. However, the pairwise correlations are low. For the most part, the EWMPCA principal components are distinct from the corresponding classical PCA principal components.

\section{Economic validation}

Has EWMPCA economic significance over and above that of the classical PCA? While there are many ways to explore this question, we focus on a particular approach. Avellaneda and Lee have demonstrated in~\cite{avellaneda-2010} that PCA can be used to generate profitable trading strategies. Can EWMPCA better them?

For each component, we obtain a trading strategy, and a backtest gives us its Sharpe ratio~\cite{sharpe-1994}. We compute the Sharpe ratios for the strategies based on the classical PCA as well as for the strategies based on EWMPCA, while keeping all parameters equal between the two methods. The results are shown in Table~\ref{tab:backtesting-results}.

\begin{table}
\centering
\begin{tabular}{ccccc}
\toprule
                    & \multicolumn{2}{c}{Data Set 1} & \multicolumn{2}{c}{Data Set 2} \\
\cmidrule{2-3} \cmidrule{4-5}
Principal component & Classical PCA & EWMPCA         & Classical PCA & EWMPCA         \\
\midrule
PC1                 & \textbf{0.65} & 0.73           & -0.32         & 0.02           \\
PC2                 & 0.43          & \textbf{1.02}  & -0.39         & -0.34          \\
PC3                 & -0.2          & 0.89           & -0.13         & \textbf{0.48}  \\
PC4                 & -0.11         & -0.11          & -0.47         & 0.26           \\
PC5                 & 0.5           & -0.33          & 0.3           & -0.39          \\
PC6                 & -0.01         & -0.13          & \textbf{0.37} & -0.16          \\
PC7                 & -0.5          & 0.04           &               &                \\
PC8                 & -0.03         & 0.06           &               &                \\
PC9                 & -0.08         & 0.23           &               &                \\
\bottomrule
\end{tabular}
\caption{Backtesting results: annualized Sharpe ratios. The maxima over the PCs are shown in bold. \label{tab:backtesting-results}}
\end{table}

We see that on both Data Set 1 and Data Set 2 Avellaneda--Lee-style statistical arbitrage strategies achieve higher maximum Sharpe ratios when used with EWMPCA as opposed to classical PCA.

\section{Implementation}

The code behind this paper is publicly available on GitHub: \url{https://github.com/sydx/xpca}

The repository contains a general-purpose Python library, \texttt{xpca.py}, and the notebooks that were used to produce the figures in this paper.

A few notes on the implementation are in order. The class \texttt{IPCA} implements the iterated PCA algorithm. It has been modelled on \texttt{sklearn.decomposition.PCA}, so that \texttt{IPCA} can be a drop-in replacement for the former. However, no attempt to achieve industrial-grade performance has been made; in particular, the functions \texttt{estimate\_eigenvalues}, \texttt{ogita\_aishima\_step}, and \texttt{ogita\_aishima} could benefit from further optimization.

The class \texttt{EWMPCA}, as the name suggests, implements the EWMPCA algorithm. It can be used in two modes (and the modes can be interleaved):
\begin{itemize}
\item the online mode, where the method \texttt{add} is applied to a single observation and returns the corresponding observation transformed to the principal component space;
\item the batch mode, where the method \texttt{add\_all} is applied to a matrix whose rows are $p$-dimensional observations; the result is, then, a matrix of principal components.
\end{itemize}

\bibliographystyle{alpha}
\newcommand{\etalchar}[1]{$^{#1}$}

\end{document}